\documentclass[twocolumns,reprint,superscriptaddress,aps,pre]{revtex4-1}
\usepackage{amsfonts}
\usepackage{amssymb}
\usepackage{amsmath}
\usepackage{upgreek}
\usepackage{bm}
\usepackage{physics}
\usepackage{graphicx}
\usepackage{gensymb,color}
\usepackage{xr}
\usepackage{natbib}
\graphicspath{ {images/} }

\def\eqS{\begin{equation}}
\def\eqE{\end{equation}}

\makeatletter
\def\@fnsymbol#1{\ensuremath{\ifcase#1 \or \dagger\or *\or \mathsection\or \mathparagraph\or \|\or **\or \dagger\dagger \or \ddagger\ddagger \else\@ctrerr\fi}}
\makeatother

\makeatletter
\newcommand*{\balancecolsandclearpage}{%
 \close@column@grid
 \cleardoublepage
 \twocolumngrid
}
\makeatother

\begin{document}
\title{Fundamental performance bounds on time-series generation using reservoir computing}
\date{\today}
\author{Daoyuan Qian}
\affiliation{Centre for Misfolding Diseases, Yusuf Hamied Department of Chemistry, University of Cambridge, Lensfield Road, Cambridge CB2 1EW, U.K.}

\author{Ila Fiete}
\thanks{fiete@mit.edu}
\affiliation{McGovern Institute for Brain Research, Massachusetts Institute of Technology, MA 02139, U.S.A.}
\affiliation{Department of Brain and Cognitive Sciences, Massachusetts Institute of Technology, MA 02139, U.S.A.}

\begin{abstract}
Reservoir computing (RC) harnesses the intrinsic dynamics of a chaotic system, called the reservoir, to perform various time-varying functions. An important use-case of RC is the generation of target temporal sequences via a trainable output-to-reservoir feedback loop. Despite the promise of RC in various domains, we lack a theory of performance bounds on RC systems. Here, we formulate an existence condition for a feedback loop that produces the target sequence. We next demonstrate that, given a sufficiently chaotic neural network reservoir, two separate factors are needed for successful training: global network stability of the target orbit, and the ability of the training algorithm to drive the system close enough to the target, which we term `reach'. By computing the training phase diagram over a range of target output amplitudes and periods, we verify that reach-limited failures depend on the training algorithm while stability-limited failures are invariant across different algorithms. We leverage dynamical mean field theory (DMFT) to provide an analytical amplitude-period bound on achievable outputs by RC networks and propose a way of enhancing algorithm reach via forgetting. The resulting mechanistic understanding of RC performance can guide the future design and deployment of reservoir networks.
\end{abstract}

\maketitle

\section{Introduction}

Random neural networks with strong connectivity exhibit chaotic dynamics \cite{Sompolinsky1988,Paz2024}. The reservoir computing (RC) paradigm seeks to functionalise the dynamics of such networks by treating the neural firing rates as a source (reservoir) of temporal basis functions which can be combined through a linear layer to produce an output of desired properties \cite{Maass2002,Maass2011,jaeger2001echo}. Time-series generation is an important and natural application of RC, where training the final output layer leads to rapid convergence of the output towards a given target function, provided that the network is stabilised by a fixed feedback loop \cite{Sussillo2009}. The appeal of RC over conventional machine learning models for time series is that while the outputs and feedback into the reservoir are optimised by training, the reservoir remains intact, so the training process can be significantly more efficient \cite{Schrauwen2007,Tanaka2019a}. Reservoirs can be instantiated as physical systems, with examples including coupled mechanical \cite{Coulombe2017} or chemical oscillators \cite{Goudarzi2013}, electronic circuits \cite{Soures2017}, photonic systems \cite{VanDerSande2017}, and quantum computers \cite{Fujii2017}. 

Typically, the readout parameters from the reservoir are trained using the recursive least-squares (RLS) method \cite{Haykin2002}. During training, either the network's current output or the target output is fed back to the system, giving rise to the first-order reduced and controlled error (FORCE) algorithm 
 \cite{Sussillo2009} and the echo state network (ESN) algorithm \cite{Jaeger2002,jaeger2001echo,Jaeger2004}, respectively. Training is successful if the output remains close to the target after the readout parameters are fixed. In the brain, it is hypothesised that the sequence-learning ability of random networks may underlie the generation of motor movement sequences in the cortex \cite{Sussillo2009}. Computational models further illustrate how low-rank perturbative feedback (from the cortex via the thalamus) to a reservoir (cortex) can stabilize and combine multiple output trajectories to generate complex motor sequences \cite{Logiaco2021}. More generally, ESNs can also act as fading memory filters and their applications are mostly found in approximating generic input/output systems with fading memory \cite{Gonon2023,Grigoryeva2018}.

While past work has focused on various applications of RC across domains, we understand little about the bounds on performance using RC and when and why training may succeed or fail in general. In \cite{Mastrogiuseppe2019}, the requirements for successful training of RC to approximate a stationary input-output function are characterised in terms of alignment between the input-output vectors, and in \cite{Susman2021} training errors for generation of sinusoidal waves by sub-chaotic linear networks are studied. For a general nonlinear network in the chaotic or sub-chaotic regime, successful learning of a target function depends on multiple factors: first (existence condition), a dynamical orbit that produces the target function must exist; second (stability condition), this orbit needs to be stable over time; and third (reach condition), the training algorithm has to be able to reach this orbit during training. In this work, we examine each condition using a combination of analytical theory and numerical simulations, leading to rules that determine whether an RC implementation of a time-series generation task can be successful. 

We find that the existence condition depends on network size and is easily satisfied for a large reservoir, but the orbit stability and the reach of the algorithm play major roles in determining which functions can be learned. To understand the latter factors, we train a network to produce sinusoidal waves of a range of amplitude $A$ and period $T$, and compute the relative training error in the $A-T$ space. We derive a stability-related phase boundary using a dynamical mean-field theory (DMFT) approach and verify its validity in simulations, and further show that a training algorithm can be tuned to achieve greater reach in the stable region. Taken together, our work illuminates distinct mechanisms that bound the performance of an RC reservoir, allowing for more informed implementation and optimisation of these systems.

\section{Random neural network reservoir computer}

\begin{figure}[htb]
\includegraphics[width=8.7cm]{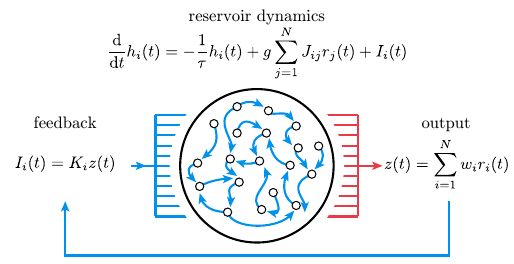}
\caption{\textbf{Implementing RC with a random neural network.} A reservoir of $N$ neurons is recurrently coupled via a random matrix $J_{ij}$. The RC output is a weighted sum of reservoir firing rates $r_{i}(t)$. A feedback loop re-injects the output into the reservoir, to push it toward a stable limit cycle. Only the output weights $w_{i}$ are trained (red lines), while the recurrent and feeback weights $K_{i}$ are fixed (blue lines).}
\label{fig:network}
\end{figure}

We take as the reservoir a network of $N$ randomly connected neurons labeled by index $i=1,2,\dots,N$. The neural membrane potential and firing rate at time $t$ are denoted $h_{i}(t)$ and $r_{i}(t)$ respectively. They are related via $r_{i}(t)=\phi[h_{i}(t)]$, where in numerical computations the non-linearity $\phi(\cdot)$ is taken to be $\phi(\cdot)\equiv\tanh(\cdot)$. The dynamical equation is
\eqS
\frac{\dd}{\dd t}h_{i}(t)=-
\frac{1}{\tau}h_{i}(t)+g\sum_{j=1}^{N}J_{ij}r_{j}(t)+I_{i}(t),
\label{eq:EOM}
\eqE
where $\tau$ is the decay time of the membrane potential of individual neruons, $J_{ij}$ is a random connectivity matrix satisfying $\expval{J_{ij}}=0$ and $ \expval{J_{ij}^{2}}=\frac{1}{N}$, $g$ sets the connection strength, and $I_{i}(t)$ is an external input into the $i$-th neuron. When $g>\frac{1}{\tau}$ and in the absence of external input, the potentials exhibit chaotic dynamics \cite{Sompolinsky1988}. The RC output $z(t)$ is the weighted sum of all reservoir neuron firing rates, and is fed back to the system via a feedback vector $K_{i}$:
\eqS
z(t)\equiv\sum_{i=1}^{N}w_{i}r_{i}(t),\quad I_{i}(t)= K_{i}z(t).
\label{eq:zI}
\eqE
The readout weights $w_{i}$ are trained while the recurrent and feedback weights $J_{ij}$ and $K_{i}$ remain fixed (Figure \ref{fig:network}). The aim of training is to find a $w_{i}$ such that $z(t)$ mimics a target sequence $f(t)$, where $f(t)$ can take complicated forms. The feedback loop defined by Eqs.~(2) can be interpreted as a rank-1 perturbation $K_{i}w_{j}$ to the random connectivity matrix $J_{ij}$, and during training this perturbation is tuned to produce the target function. The ideal $K_{i}w_{j}$ is in general correlated to the instantiation of $J_{ij}$ instead of mimicking a random perturbation \cite{Schuessler2020,Susman2021}, and the perturbation can also be set by hand to produce desired functions \cite{Logiaco2021} instead of using a training algorithm.

\section{Existence of a solution}

Least-squared minimisation of the integrated quadratic error between $z(t)$ and $f(t)$ leads to an expression for the weight vector required to solve the task \cite{Sussillo2009}:
\eqS
w_{i}=\sum_{j=1}^{N}P_{ij}\int r_{j}(t)f(t)\dd t,
\label{eq:idealW}
\eqE
where $P_{ij}$ is the regularised inverse coorelation matrix
\eqS
P_{ij}\equiv\left(R_{ij}+\frac{1}{\alpha}\delta_{ij}\right)^{-1},\quad R_{ij}\equiv\int r_{i}(t)r_{j}(t)\dd t.
\label{eq:PR}
\eqE
The regularizer $\alpha$ is added to avoid numerical issues with matrix inversion. Substituting Eq.~\eqref{eq:idealW} into the first of Eq.~\eqref{eq:zI} gives the output of the RC network as
\eqS
z(t)=(D*f)(t),\quad D(t,t')=\sum_{i,j=1}^{N}r_{i}(t)P_{ij}r_{j}(t'),
\label{eq:Dorigin}
\eqE
where $*$ denotes a convolution. Thus, the existence of an orbit in the RC dynamics that reproduces the target requires that $D(t,t')\approx\delta(t-t')$. For a network without feedback that evolves freely, and when the regulariser is absent, $D(t,t')$ behaves as a normalised Gaussian with effective variable $t-t'$ if the time horizon $T$ (the length of the target sequence) is short enough, $T<N\tau_{0}$, where $\tau_{0}$ is the auto-correlation time of a single neuron firing rate (Appendix A1). When $R_{ij}$ is full-rank, we can remove the regulariser from Eq.~\eqref{eq:PR} and the peak takes the value $\expval{D(t,t)}\approx\frac{N}{T}$, where the average is taken over $t$. However, if $T$ is too short (compared to $N\tau_{1}$ where $\tau_{1}$ is a de-correlation timescale, Appendix A1), $R_{ij}$ becomes low-rank and a regulariser is needed to compute $P_{ij}$. With the regulariser, the peak in $D(t,t')$ is now (Appendix A2)
\eqS
\expval{D(t,t)}\approx\frac{Np}{T}, 
\label{eq:DTT}
\eqE
where $p$ is the fraction of eigenvalues of $R_{ij}$ significantly larger than $\frac{1}{\alpha}$. The peak width of $D(t,t')$ is thus $\frac{T}{Np}$. Going back to the RC setting where a feedback loop is present to generate a target output, $Np$ depends on the reservoir parameters, the regularizer, as well as the target function. For a good RC fit to a target, we would ideally have the width $\frac{T}{Np}$ to be much smaller than the fitting time horizon $T\gg\frac{T}{Np}$, so the condition for a successful RC implementation is that the relative smearing width is much less than 1, $1/Np\ll 1$. Furthermore, if we take $1/Np$ as a proxy for the fitting error denoted by $\eta$, we arrive at a scaling law for $\eta$ with $N$, assuming $p$ is constant:
\eqS
\eta\sim\frac{1}{Np}\propto\frac{1}{N}.
\label{eq:etaN}
\eqE
This scaling is empirically observed in numerical simulations of FORCE learning at different $N$, in the region of optimal training (Appendix B). In sum, the existence condition can be satisfied by sufficiently large reservoirs, and the analysis below assume the large-$N$ limit.

\section{Stability condition from DMFT}

After the weight vector $w_{i}$ has been learned and weight updates have stopped, the network orbit must be stable. To analytically derive a stability condition for the dynamics, we set $f(t)$ to be a sinusoidal wave of amplitude $A$, period $T$, and angular frequency $\omega\equiv\frac{2\pi}{T}$:
\eqS
f(t)=A\sin\omega t.
\label{eq:sin}
\eqE
Training a reservoir to produce a sinusoidal wave has been used in the past to study sub-chaotic linear networks \cite{Susman2021} or in a narrow parameter range \cite{jaeger2001echo} but a more general investigation is lacking. Letting $K_{i}=1$ for now, by DMFT the mean membrane potential $\bar{h}(t)$ evolves as $\frac{\dd}{\dd t}\bar{h}(t)=-\frac{1}{\tau}\bar{h}(t)+A\sin\omega t$ \cite{Rajan2010a}, so that 
\eqS
\bar{h}(t)=\frac{A}{\sqrt{1/\tau^{2}+\omega^{2}}}\sin(\omega t+\theta),
\label{eq:hbar}
\eqE
for some phase factor $\theta$. When the network is perturbed from the ideal orbit by $\delta h_{i}(t)$, the equation governing its evolution is given by the perturbation of Eq.~\eqref{eq:EOM}
\eqS
\frac{\dd}{\dd t}\delta h_{i}(t)=-\frac{1}{\tau}\delta h_{i}(t)+g\sum_{j}J_{ij}\left[1-r^{2}_{j}(t)\right]\delta h_{j}(t).
\label{eq:perturb}
\eqE
Recall that $r_{j}(t) \in [-1,1]$, the eigenvalues of $J_{ij}$ have a maximum magnitude of approximately 1, and $g>\frac{1}{\tau}$ for a chaotic reservoir. Exponential growth of $\delta h_{j}(t)$ can be expected when $r_{j}(t)\approx0$, and the perturbation shrinks when $r_{j}(t)\approx\pm1$. The boundary between a stable and an unstable network is when the growth and decay of $\delta h_{i}(t)$ over one period exactly cancel each other. To quantify this statement, we make the mean-field replacement 
\eqS
r_{j}(t)\approx\tanh\bar{h}(t),
\label{eq:0order}
\eqE
and since the initial perturbation $\delta h_{i}(0)$ is arbitrary, we focus on the eigenvector of $J_{ij}$ with the largest real eigenvalue, of order 1. The reduced model is then, upon re-arrangement and dropping the subscript for brevity,
\eqS
\frac{\dd}{\dd t}\ln\delta h(t)=-\frac{1}{\tau}+g\left[1-\tanh^{2}\bar{h}(t)\right].
\label{eq:reduced2}
\eqE
Network stability is given by the time integral of the right hand side, with positive and negative values indicating unstable and stable orbits respectively. Since $\bar{h}$ is given explicitly by Eq.~\eqref{eq:hbar}, the instability boundary satisfies
\eqS
\int_{-\infty}^{\infty}\dd t'\left\{-\frac{1}{\tau}+g\left[1-\tanh^{2}\left(\frac{A\sin t'}{\sqrt{1/\tau^{2}+\omega^{2}}}\right)\right]\right\}=0.
\label{eq:inttozero}
\eqE
In the above we have re-scaled and shifted the dummy time variable by $t'\equiv\omega t+\theta$ such that the parameters $A$ and $\omega$ appear in a combined form $A/\sqrt{1/\tau^{2}+\omega^{2}}$. As a result, when both $\tau$ and $g$ are fixed, the stable-unstable boundary on an $A-\omega$ plot should follow $A/\sqrt{1/\tau^{2}+\omega^{2}}=\text{const}$, or
\eqS
A\propto\sqrt{\frac{1}{\tau^{2}}+\omega^{2}}.
\label{eq:phaseboundary}
\eqE
Because $T=2\pi/\omega$, we expect the success/failure boundary in $A-T$ space to have the scaling $A\propto T^{-1}$ when the period $T$ is much smaller than $\tau$, and plateaus when $T$ becomes larger than $\tau$. If $f(t)$ is composed of a few sinusoidal waves, the mean potential Eq.~\eqref{eq:hbar} acquires additional terms of the form $A/\sqrt{1/\tau^{2}+n^{2}\omega^{2}}$ for some integer $n$ (Appendix C). The $T^{-1}$ scaling still holds for small $T$. Interestingly, Eq.~\eqref{eq:phaseboundary} implies that in the sub-chaotic regime of $\frac{1}{\tau}>g$, all orbits should be stable.


\section{Reach of the learning algorithm}

\begin{figure*}[htb]
\includegraphics[width=17.8cm]{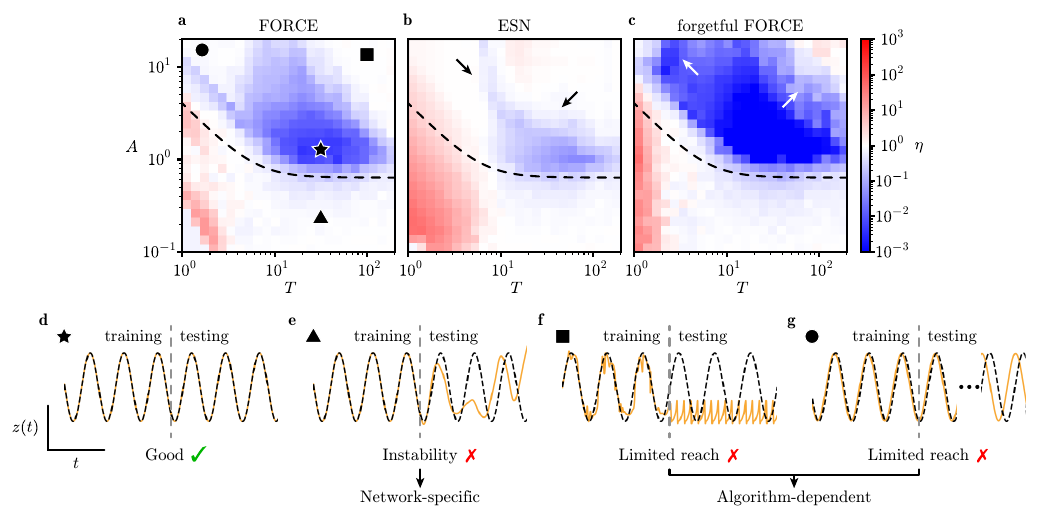}
\caption{\textbf{Probing performance bounds on RC by training on sinusoidal waves.} We compare training phase diagrams of three different algorithms: \textbf{a} FORCE learning from \cite{Sussillo2009}, \textbf{b} ESN learning, and \textbf{c} learning with a forgetting parameter $\gamma=0.002$. Black dashed lines are instability-induced failure phase boundaries from Eq.~\eqref{eq:phaseboundary}, and the dark blue (success) regions do not extend beyond this boundary in all three cases. ESN is less expressive (black arrows indicating shrinkage of the blue region), while forgetful FORCE enlarges the blue region (white arrows). Examining the output traces from FORCE simulations show different types of failure. In \textbf{d}-\textbf{g}, orange lines are the network output and black bashed lines are the target output. \textbf{d} In the success region, the output and target overlap during and after training. \textbf{e} If the target function is outside the stable region, the output matches the target during training but drifts away from it during testing. \textbf{f}, \textbf{g} Within the stable regime, the network can still fail to learn the target function due to limited algorithm reach, and these failures are characterised by output-target mismatches both during and after training.}
\label{fig:trainings}
\end{figure*}

The existence of a stable solution orbit does not necessarily imply that a learning algorithm can reach it. To understand the learning algorithms, we first outline the RLS update scheme \cite{Haykin2002}, used in both FORCE and ESN \cite{Sussillo2009,Jaeger2002}. The weight vector $w_{i}$ and inverse correlation matrix $P_{ij}$ are kept in memory and updated at time intervals $\Delta t$. Using the shorthand $\bm{P}_{0}\equiv P_{ij}(0)$, $\bm{w}_{0}\equiv w_{j}(0)$, $\bm{P}\equiv P_{ij}(t)$, $\bm{w}\equiv w_{j}(t)$, $\bm{r}\equiv r_{j}(t)$, $\bm{P}'\equiv P_{ij}(t+\Delta t)$, $\bm{w}'\equiv w_{j}(t+\Delta t)$, and $\bm{r}'\equiv r_{j}(t+\Delta t)$, the RLS algorithm is
\eqS
\begin{split}
  \bm{P}_{0}=&\alpha\bm{1},\quad \bm{w}_{0}=\bm{0},\\
  \bm{P}'=&\bm{P}-\frac{\bm{P}\bm{r}'\bm{r}^{\prime T}\bm{P}}{1+\bm{r}^{\prime T}\bm{P}\bm{r}'},\\
  \bm{w}'=&\bm{w}-\bm{P}'\bm{r}'e',
\end{split}
\label{eq:update}
\eqE
where $e'\equiv z(t+\Delta t)-f(t+\Delta t)$ is the instantaneous error at $t+\Delta t$. In essence, the RLS algorithm performs the integrals in Eqs.~\eqref{eq:idealW} and \eqref{eq:PR} on the fly. The difference between FORCE and ESN is that during training in FORCE learning, the feedback is set to $I_{i}(t)=K_{i}z(t)$, while in ESN it is $I_{i}(t)=K_{i}f(t)$. Feeding the network its own output allows system fluctuations to be corrected during training, leading to better performance in finding the ideal orbit \cite{Sussillo2009,jaeger2001echo}. The time-evolution of the error can be written using Eq.~\eqref{eq:update} as
\eqS
e'-e=-\bm{r}^{\prime T}\bm{P}'\bm{r}'e'+(\bm{w}\Delta\bm{r}-\Delta f),
\label{eq:osci}
\eqE
with $e\equiv z(t)-f(t)$, $\Delta\bm{r}\equiv\bm{r}(t+\Delta t)-\bm{r}(t)$, and $\Delta f\equiv f(t+\Delta t)-f(t)$. The first term on the right hand side of Eq.~\eqref{eq:osci} represents an exponential decay of the error, while the next two terms represents the new error generated via the mismatch between the rates of change of the target and actual outputs. The exponential decay needs to suppress the oscillatory drive represented by $\bm{w}\Delta\bm{r}-\Delta f$ during training to achieve a successful reproduction of the target function, so the expectation value of $\bm{r}^{\prime T}\bm{P}'\bm{r}'$ given by Eq.~\eqref{eq:DTT} can be interpreted as the expressivity of the network. 
It is then easy to compare the ESN and FORCE training algorithms: ESN feeds the network with the target function with no fluctuation, while FORCE feeds the network with fluctuations that introduce larger eigenvalues to the correlation matrix. As a result, the $\bm{r}^{\prime T}\bm{P}'\bm{r}'$ value is larger for FORCE than for ESN, leading to greater algorithmic reach. We also propose a way to increase the reach of FORCE further by gradually increasing the $\bm{r}^{\prime T}\bm{P}'\bm{r}'$ term by introducing an intermediate matrix $\bm{P}''$ at each update step
\eqS
\begin{split}
\bm{P}''=&\frac{1}{1-\gamma}\bm{P},\\
\bm{P}'=&\bm{P}''-\frac{\bm{P}''\bm{r}'\bm{r}^{\prime T}\bm{P}''}{1+\bm{r}^{\prime T}\bm{P}''\bm{r}'},
\end{split}
\eqE
with a `forgetting' parameter $\gamma$. This is equivalent to multiplying the correlation matrix $R_{ij}$ by a factor $(1-\gamma)$ before each update, and the system retains a memory of its past only over the timescale $\Delta t/\gamma$. This algorithm can then be interpreted as a `forgetful' version of FORCE. Forgetting should improve algorithm reach in regions where a stable solution exists, but should not affect instability-induced failure. 

\section{Numerical verifications}

We test the performance of FORCE, ESN, and forgetful FORCE algorithms by training a random neural network to produce the sinusoidal wave as in Eq.~\eqref{eq:sin}. We use typical RC parameters \cite{Sussillo2009} $\tau=1$, $g=1.5$, $N=1000$, $\Delta t=0.1$, $\alpha=1$, an integration time step $\dd t=0.01$, and draw $K_{i}$ uniformly and independently from the interval $(-1,1)$. The training phase is simulated for 400 time units, after which a testing phase ensues with the weight $w_{i}$ fixed for another 400 time units. Simulating the training phase with 200 or 600 time units does not show significant differences to 400 time units. We perform a parameter sweep over $A$ and $T$, and 20 different network realisations are simulated at each parameter combination. We quantify the success or failure of a training episode by computing the relative error $\eta$ in the testing phase with fixed $w_{i}$
\eqS
\eta\equiv\sqrt{\expval{\left[f(t)-z(t)\right]^{2}}}/A,
\label{eq:simEta}
\eqE
where the average is taken over time $t$. $\eta$ is then equivalent to the noise-to-signal ratio, and successful training should lead to $\eta\ll1$. 

Computing the average $\eta$ for a range of $A$ and $T$ for all three training algorithms indeed confirms our results pertaining to stability and reach. In the low-$T$ (high-$\omega$) regime, the FORCE boundary exhibits the predicted scaling relation $A\propto T^{-1}$ that plateaus at higher $T$ (Figure \ref{fig:trainings}a, dashed black line). Regardless of learning rule --- FORCE (Figure \ref{fig:trainings}a), ESN (Figure \ref{fig:trainings}b), or forgetful FORCE (Figure \ref{fig:trainings}c) --- the empirical success region does not extend below this stability boundary. However, the algorithms differ from each other in the high-$A$, stable region. For the same reservoir parameter $g$, ESN performs worse than FORCE since the feedback does not include network fluctuations \cite{jaeger2001echo,Sussillo2009}, while forgetful FORCE has highest reach because it keeps only relevant correlation information and anneals away the regulariser. Examining network outputs for individual runs shows qualitative differences between instability and low-reach failures. For a successful training run, the output matches the target both during and after training (Figure \ref{fig:trainings}d). In the unstable region, the output matches the target during training but diverges from it when weight updates are stopped (Figure \ref{fig:trainings}e). In the low-reach region, the output differs from the target both during and after training: the algorithm fails to reach the ideal orbit even during training (Figure \ref{fig:trainings}f, \ref{fig:trainings}g).

\begin{figure}[htb]
\includegraphics[width=8.7cm]{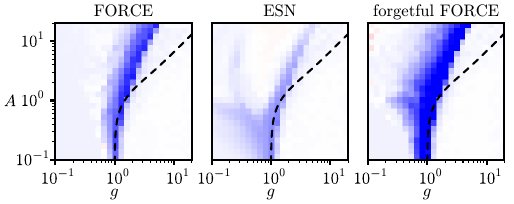}
\caption{\textbf{$\bm{A-g}$ parameter sweep reveals instability boundary near $\bm{g=1}$.} Analytical DMFT result suggests the network should always be stable to the left of the black dashed lines. Simulations (blue) indeed show a sharp instability boundary that coincides at low amplitudes (boundary goes to $A\to0$ with $g\to1$). The deviation between the two at large $g$ indicates a breakdown of DMFT due to the increased effect of fluctuations. In the stable regime, training can still fail in an algorithm-dependent manner, indicating varying degrees of training success among the three tested algorithms. The colour code follows that of Figure \ref{fig:trainings}a - \ref{fig:trainings}c.}
\label{fig:gscan}
\end{figure}

The DMFT expression Eq.~\eqref{eq:inttozero} also suggests that the network should always be stable in the subchaotic regime, $g<\frac{1}{\tau}=1$. This observation is also made in \cite{Rivkind2017} by studying the open-loop gain of the network. However, sub-chaotic reservoirs are documented to have worse performance in sequence learning \cite{Sussillo2009}, but the reasons are unclear. To gain more insight on how performance depends on $g$, we perform a parameter sweep of $A$ against $g$, with $T=30$ and all other parameters the same as before. All three algorithms exhibit a sharp success/failure boundary (Figure \ref{fig:gscan}). For a significant region, this boundary is near $g=1$, confirming our expectation that the chaotic regime is detrimental to network stability. We also numerically solved the analytical expression from Eq.~\eqref{eq:inttozero} to compute the $A-g$ phase boundary (Figure \ref{fig:gscan}, black dashed lines). Despite the deviation of the analytical $A-g$ curve from simulations for $g$ away from criticality, they agree qualitatively. The deviations at large $g$ are possibly due to the de-stabilising effect of fluctuations omitted in the DMFT analysis. In contrast, the $A-T$ boundary Eq.~\eqref{eq:phaseboundary} has greater applicability than we had expected, indicating First-order fluctuations in $r_{j}(t)$ can also depend on $A$ and $T$ in a combined form $A/\sqrt{1/\tau^{2}+\omega^{2}}$. Future work could use a first-order approximation of $r_{j}(t)$ in addition to Eq.~\eqref{eq:0order}, although the analytical intractability of typical DMFT calculations means a simple closed-form solution is unlikely, so we omit this in the current work. For $g<1$, although the network should be stable at all $A$, the degree of training success varies among the algorithms (Figure \ref{fig:gscan}), indicating that algorithm reach plays a more important role for sub-chaotic networks than the detailed value of $g$ in this regime. Notably, algorithm reach deteriorates with decreasing $g$, and the best training quality can be achieved at $g\approx1$, at the edge of chaos.


\section{Conclusion}

In this work, we delineated two distinct factors that are required to successfully train a reservoir computing network with a large reservoir: dynamic stability and learning algorithm reach. The former is an intrinsic property of the network and can be understood using a DMFT approach. We demonstrate the effect of the latter via direct computation of training phase diagrams using three different training algorithms. The new theoretical understanding of RC also carries practical implications. We proposed a new training algorithm whose reach improves on that of existing algorithms by only taking into consideration recent trajectory data and with annealing of the regularisation term. Further, applying different training algorithms to a single system can potentially illuminate an algorithm-independent performance boundary, such that if the desired function falls outside this boundary then it is necessary to modify the reservoir itself. Making biological reservoirs using cultured neurons is also an area under active exploration \cite{Lindell2024} and experimentally measuring the stability boundary could provide a way to characterise an \textit{in vitro} network using coarse-grained parameters that are typically employed in computational models. Separating network-specific and algorithm-dependent failures can be a new approach to study experimentally constructed RC systems and facilitate their development.
\\

\textbf{Conflict of interest} The authors declare no conflict of interest.
\\

\textbf{Code availability} The simulation scripts used in this work will be made available as a GitHub repository upon acceptance for publication.
\\

\textbf{Acknowledgement} The study is funded by Transition Bio Ltd (D.Q.). IRF is supported in part by the Simons Foundation (SCGB program 1181110), the ONR (award N00014-19-1-2584), and the NSF (CISE award  IIS-2151077 under the Robust Intelligence program). The authors thank Mr.~David Clark, Dr.~Sarthak Chandra, and Dr.~Laureline Logiaco for insightful discussions and comments on the manuscript.
\\

\balancecolsandclearpage

\appendix

\section{Properties of the correlator $\bm{D(t,t')}$}

In this section, we use the summation convention where repeated indices are assumed to be summed unless stated otherwise.
\\

\subsection{Full rank random dynamics}

In the absence of input currents, the firing rates $r_{i}(t)$ have no periodic structure. We are interested in the function
\eqS
D(t,t')=r_{i}(t)P_{ij}r_{j}(t')
\eqE
where $P_{ij}$ is the inverse matrix of the integrated correlation
\eqS
P_{ij}\int r_{j}(t)r_{k}(t)\dd t=\delta_{ik}.
\eqE
We have omitted the regularizer for now. Integrating $D(t,t)$ over $t$ leads to
\eqS
\int D(t,t)\dd t=P_{ij}R_{ij}=\delta_{ii}=N,\quad\expval{D(t,t)}=\frac{N}{T}.
\label{eq:S_NT}
\eqE
This is an exact result, and deviations from this can arise due to numerical issues with a large $N$ or a small $T$, associated with an $R_{ij}$ matrix of rank less than $N$. In particular, we can view $\{r_{i}(t)| 0<t<T\} $ as a random sequence with a decorrelation time-scale of $\tau_{1}$, so effectively as a list of $\frac{T}{\tau_{1}}$ random numbers. $\tau_{1}$ is not a network-specific timescale, as the decorrelation time-scale depends on the simulation hardware's ability to distinguish changes in $r_{i}(t)$, and thus only a numerical artefact. The full-rank condition for $R_{ij}$ reads
\eqS
N\lesssim\frac{T}{\tau_{1}}.
\eqE
Eq.~\eqref{eq:S_NT} gives the peak height of $D(t-t')\equiv D(t,t')$. To provide intuition on why $D(t-t')$ should resemble a Dirac delta, we write $r_{i}(t)$ as a rectangular data matrix $\bm{x}$ of dimension $\frac{T}{\dd t}\cross N$, so $D(t,t')$ can be written as $\bm{D}=\bm{x}(\bm{x}^{T}\bm{x})^{-1}\bm{x}^{T}$. Furthermore, for a constant $y$
\eqS
\begin{split}
(\bm{x}^{T}\bm{x}+y\bm{I})\bm{x}^{T}=&\bm{x}^{T}(\bm{x}\bm{x}^{T}+y\bm{I})\\
\bm{x}^{T}(\bm{x}\bm{x}^{T}+y\bm{I})^{-1}=&(\bm{x}^{T}\bm{x}+y\bm{I})^{-1}\bm{x}^{T},
\end{split}
\label{eq:S_inv}
\eqE
and in the limit of $y\to0^{+}$ these are the Moore-Penrose inverse of $\bm{x}$, denoted as $\bm{x}^{+}$ \cite{Penrose1955}:
\eqS
\bm{x}^{+}=\lim_{y\to0^{+}}\bm{x}^{T}(\bm{x}\bm{x}^{T}+y\bm{I})^{-1}=\lim_{y\to0^{+}}(\bm{x}^{T}\bm{x}+y\bm{I})^{-1}\bm{x}^{T},
\eqE
leading to
\eqS
\begin{split}
\bm{D}=&\bm{x}(\bm{x}^{T}\bm{x})^{-1}\bm{x}^{T}=\bm{x}\bm{x}^{+}\approx\bm{I},
\end{split}
\eqE
where $\bm{I}$ in the last part is the identity matrix of dimension $\frac{T}{\dd t}\cross\frac{T}{\dd t}$, which is a Dirac delta in the continuous picture. The approximation becomes equality if $\bm{x}$ is invertible, however in practice the peak has a finite width. The mechanism becomes clear if we perform singular value decomposition on $\bm{x}$ by writing $\bm{x}=\bm{U\Sigma V}^{T}$ and $\bm{x}^{+}=\bm{V\Sigma}^{+}\bm{U}^{T}$, where $\bm{\Sigma}=\text{diag}(\lambda_{1},\lambda_{2},\dots,\lambda_{N},0,\dots,0)$ with $\frac{T}{\dd t}-N$ zeros, and $\bm{\Sigma}^{+}=\text{diag}(1/\lambda_{1},1/\lambda_{2},\dots,1/\lambda_{N},0,\dots,0)$. Then $\bm{xx}^{+}=\bm{U}\bm{I}^{+}\bm{U}^{T}$, with $\bm{I}^{+}$ comprising $N$ 1's on the diagonal and the rest are 0. As a result, when $\frac{T}{\dd t}-N$ becomes large, more parts of $\bm{U}^{T}$ are masked-out and the quality of the Dirac delta worsens. This continues until in the large-$T$ regime, we can instead assume the neurons behave as independent units such that
\eqS
\expval{r_{i}(t)r_{j}(t')}=\delta_{ij}\expval{r^{2}}C(t-t'),
\eqE
for a constant $\expval{r^{2}}$ and a normalised single-site correlation satisfying $C(0)=1$. It is straightforward to show, in this limit,
\eqS
D(t-t')=\frac{N}{T}C(t-t').
\label{eq:corrLargeT}
\eqE
The integral over $(t-t')$ is now not normalised to 1 anymore but instead $N\tau_{0}/T$ for the single-site correlation time $\tau_{0}$. The transition occurs at $N\tau_{0}/T\approx1$, or $T\approx N\tau_{0}$.
\\

\begin{figure}[htb]
\includegraphics[width=8.7cm]{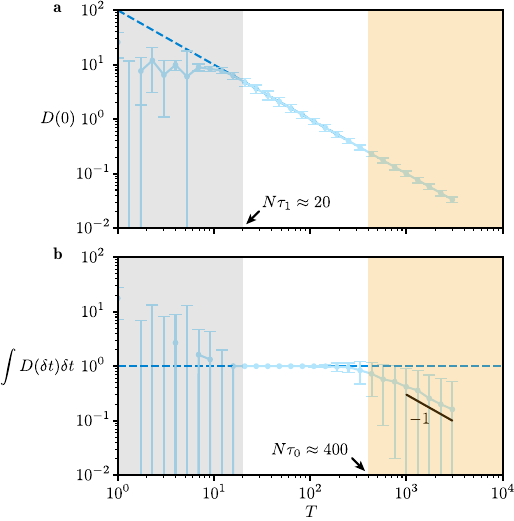}
\caption{\textbf{Properties of $\bm{D(t-t')}$}. \textbf{a} $D(0)$ computed using $N=100$ neurons (blue solid line) agrees with $\frac{N}{T}$ (blue dashed line) for $T>20$, giving $\tau_{1}\approx0.2$. This is the timescale below which inversion of $R_{ij}$ becomes numerically unstable (grey region). \textbf{b} Integral of $D(t-t')$ stays 1 for $N\tau_{1}<T<N\tau_{0}$ where $\tau_{0}\approx4$. The independent neuron assumption becomes valid for large $T$ (orange region), as indicated by the $T^{-1}$ scaling (black solid line).}
\label{fig:S_D}
\end{figure}

\begin{figure}[htb]
\includegraphics[width=8.7cm]{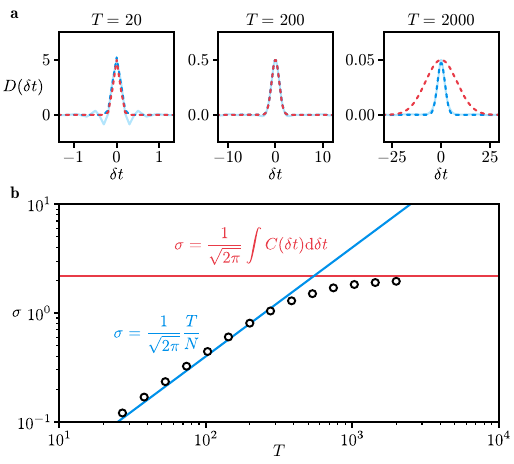}
\caption{\textbf{Explicit computation of $\bm{D(t-t')}$.} \textbf{a} $D(t-t)$ computed at a range of $T$ (light blue solid lines) give good fits to Gaussians (dark blue dashed lines). The expected peak height is $\frac{N}{T}$, and we plot normalised Gaussians with the same peak value too (red dashed lines). At $T=20$ and 200 (left and mid), integrals of $D(t-t')$ are normalised to 1 since the Gaussian fits coincide with the normalised Gaussian, but at large $T$ the integral of $D(t-t')$ becomes smaller due to de-correlation. \textbf{b} Plotting the standard deviation of the fitted Gaussian (scatter points) shows the expected scaling $\sigma\propto T$ (blue solid line) at small $T$, but plateaus at higher $T$ (red solid line).}
\label{fig:Corr}
\end{figure}

To test these results, we perform simulations of a free random network with 1000 neurones, and select $N=100$ neurons to compute the $D(t-t')$ correlator. 1000 neurons are used in the actual simulation to produce chaotic activity. The peak height $D(0)$ agrees with Eq.~\eqref{eq:S_NT} for $T>N\tau_{1}\approx20$ (Figure \ref{fig:S_D}a). The integral $\int D(\delta t)\dd\delta t$ stays at 1 for $T<N\tau_{0}$ where $\tau_{0}\approx4$, beyond which the system resembles a reservoir of independent neurons (Figure \ref{fig:S_D}b). The $\tau_{0}\approx4$ also agrees with the typical correlation time of a random neural network \cite{Sompolinsky1988,Rajan2010a}. Examining individual $D(t-t')$ forms further shows that for small $T=20$ and 200, $D(t-t')$ are approximately normalised Gaussians as expected (Figure \ref{fig:Corr}a, left and mid), and becomes un-normalised for larger $T$ (Figure \ref{fig:Corr}a, right). The standard deviation $\sigma$ of the fitted Gaussian follows the theoretical predictions in both the small-$T$ and large-$T$ regime (Figure \ref{fig:Corr}b), where for large-$T$ we explicitly compute $\int C(\delta t)\dd\delta t$ and use Eq.~\eqref{eq:corrLargeT} to get $\sigma\approx\frac{1}{\sqrt{2\pi}}\int C(\delta t)\dd\delta t$.

\begin{figure*}[htb]
\includegraphics[width=17.8cm]{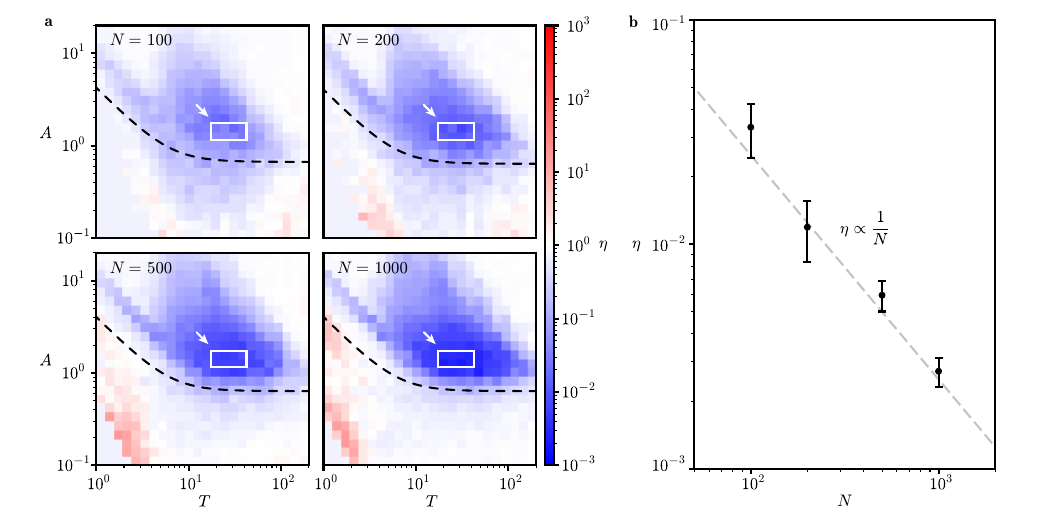}
\caption{\textbf{Optimal error scaling with $N$.} \textbf{a} Computing FORCE learning phase diagrams for various $N$ shows an increase in the relative error $\eta$ with decreasing $N$, and a small $A/T$ window (white rectangles and arrows) is used to compute the minimum $\eta$ for each $N$. \textbf{b} Plotting $\eta$ averaged over the selected region as a function of $N$ shows a $\frac{1}{N}$ scaling.}
\label{fig:trainingsN}
\end{figure*}

\subsection{Correlator with a regularizer}

For typical applications the network size $N$ is large compared to the time horizon $T$ normalised by $\tau_{0}$, $N\gg T/\tau_{0}$. In the stability study, we use $N=1000$ and $T<200$. In this regime, we expect $D(t,t+\delta t)$ to be normalised Gaussians. The practical problem with a large $N$ relative to $T$ is inversion of the correlation matrix $R_{ij}$. Assume that $R_{ij}$ has $Np$ eigenvalues significantly larger than $\frac{1}{\alpha}$ and $N(1-p)$ eigenvalues significantly smaller than $\frac{1}{\alpha}$. These $N(1-p)$ small eigenvalues become approximately $\frac{1}{\alpha}$ when the regulariser is added, so that the inverse matrix $P_{ij}$ has $Np$ eigenvalues near 0 (which become 1 when multiplied by the large eigenvalues of $R_{ij}$) and $N(1-p)$ eigenvalues at $\alpha$ (which vanish when multipled by the small eigenvalues of $R_{ij}$). As a result,
\eqS
\int D(t,t)\dd t=Np.
\eqE

\section{Relative error scaling with $\bm{N}$}

To test Eq.~\eqref{eq:etaN} on a qualitative level, we perform FORCE learning as described in section VI with $N=100$, 200, 500, and 1000, keeping all other parameters the same. The training phase diagrams show that with decreasing $N$, the success region expands and crosses the instability boundary (Figure \ref{fig:trainingsN}a), indicative of a breakdown of the DMFT result. Near the region of $T\approx25$ and $A\approx1.5$ the relative error $\eta$ is observed to be the lowest (Figure \ref{fig:trainingsN}a, white rectangles and arrows), and we hypothesise that the $\eta$ here is most likely to be limited by the finite resolution of the network to learn the target function. We compute $\eta$ averaged over this small window and plot it as a function of $N$, and the $\frac{1}{N}$ scaling is observed as predicted by Eq.~\eqref{eq:etaN}. Note that the $\eta$ here is computed according to Eq.~\eqref{eq:simEta}, and the recovered scaling indicates Eq.~\eqref{eq:simEta} is closely related to Eq.~\eqref{eq:etaN}.

\section{Learning a complicated wave}

Suppose the target function to be learned is of the form
\eqS
f(t)=A\sum_{n=1}^{\infty}a_{n}\sin(n\omega t),
\eqE
for some parameter set $a_{n}$, and $A$ now controls the overall amplitude. The mean membrane potential $\bar{h}(t)$ is given by the solution of
\eqS
\frac{\dd}{\dd t}\bar{h}(t)=\frac{1}{\tau}\bar{h}(t)+f(t),
\eqE
and since the equation is linear in $\bar{h}(t)$ we simply get 
\eqS
\bar{h}(t)=A\sum_{n=1}^{\infty}\frac{a_{n}}{\sqrt{1/\tau^{2}+n^{2}\omega^{2}}}\sin(n\omega t+\theta_{n}).
\eqE
When $T$ is small ($\omega$ is large) compared to $\tau$, the $1/\tau^{2}$ term becomes negligible so we have
\eqS
\begin{split}
\bar{h}(t)&\approx A\sum_{n=1}^{\infty}\frac{a_{n}}{n\omega}\sin(n\omega t+\theta_{n})\\
&=AT\sum_{n=1}^{\infty}\frac{a_{n}}{2\pi n}\sin(nt'+\theta_{n}),
\end{split}
\eqE
where $t'\equiv\omega t$ is the dummy integration variable in Eq.~\eqref{eq:inttozero}, and the prefactor $AT$ leads to the overall $A\propto\frac{1}{T}$ scaling.

\bibliographystyle{ieeetr}
\bibliography{library.bib}{}

\end{document}